**Reply to the correspondence: "On the fracture toughness of bioinspired ceramic materials"**

Florian Bouville [1,2*], Eric Maire [2], Sylvain Meille[2], Bertrand Van de Moortèle [3♦], Adam J. Stevenson [1], Sylvain Deville [1]

[1] Ceramic Synthesis and Functionalisation Lab, UMR3080 CNRS/Saint-Gobain, Cavaillon, France

[2] Université de Lyon, INSA-Lyon, MATEIS CNRS UMR5510, Villeurbanne, France

[3] Laboratoire de Géologie de Lyon, Ecole Normale Supérieure de Lyon, Lyon, France

* now at: Complex Materials, Department of Materials, ETH Zürich 8093 Zürich, Switzerland

♦ passed away on January 2, 2015

With very few exceptions, the most common approach to increase the toughness of ceramic materials is the introduction of a ductile phase, able to dissipate more energy than brittle materials. However, the presence of a ductile phase systematically comes at the expense of strength [1,2]. We recently demonstrated in Bouville *et al.* [3] a different approach, where a high toughness (we claimed to measure up to 22 MPa.m$^{1/2}$) was obtained in composite materials made of brittle components only (glass and ceramics) with a brick and mortar architecture. Because of the absence of any ductile phase, these materials retain the strength of their main component (470 MPa). The R-curves, which quantify how the toughness increases when the crack propagates, were measured following published guidelines [2].

In these studies of bioinspired materials, toughness is obtained by a combination of several toughening mechanisms, each of them resulting from structural features developed at different size scales. This multiscale approach renders the evaluation of the toughness essential, albeit more difficult. The shape of the stress-strain curves for the single edge notched beam (SENB) samples (Figure 1) and thus of the R-curve depends, to some extent, on the nature of the extrinsic toughening mechanisms. Materials with similar compositions but a different microstructure have been shown to exhibit R-curves with very different shapes, with either a quasi-linear toughness increase with crack length (See Fig 6a of [4], or [3,5]), or a quick rise followed by a plateau [2,4,6], traditionally observed in materials where crack bridging occurs.

An additional complexity results from the sample size effect, as pointed out in the correspondence of Prof. Ritchie. Such effects have been known for a long time and must of course be taken into account. When the crack length is too long compared to the uncracked ligament length, the R-curve adopts an upward concave shape. The determination of the transition can be difficult, in particular when only a limited number of data points in the R curve are found. This transition is also more difficult to observe when the R-curve initially has a downward concave shape.

Several criteria have been proposed to determine the maximum crack extension still valid for toughness measurements. Although ASTM standards were claimed to "provide a somewhat arbitrary conservative limit" [4], more recent papers have acknowledged that data beyond the ASTM limit are not valid [6]. The ASTM standards are usually much more conservative. Another less conservative approach was proposed [4], based on the J-integral. In his correspondence, Prof Ritchie proposed a third criterion, based on the presence of a concave upward increase of the R-curve. This last criterion actually corresponds to that defined in the ASTM standard: the maximum valid crack extension corresponds to the limit where the toughness is not geometry dependent. Depending on the validity criterion chosen (J-integral or ASTM), the reported toughness can greatly vary, decreasing in some studies from 19 to 12 MPa.m$^{1/2}$ [7], or from 16 to 7 MPa.m$^{1/2}$ [2,4].

We followed the J-integral criteria (defined in Prof. Ritchie's previous work [2]) as the crack extension where the total energy J starts increasing drastically. We thus claimed a maximum value of $K_J$ at large extension of 22 MPa.m$^{1/2}$ at room temperature and 21 MPa.m$^{1/2}$ at 600°C, which corresponds to a crack extension of 0.45 to 0.5 mm (small variations were found between the samples). The average value of toughness measured using the ASTM standards is 17.3 MPa.m$^{1/2}$. If instead of the ASTM standard we follow the guidelines proposed by Prof Ritchie, the measured toughness values are in the 15-19 MPa.m$^{1/2}$ range, with an average value of 16.6 MPa.m$^{1/2}$ (Figure 2), thus close to that obtained following the ASTM standard. The crack length at which the inflexion point is observed is very close to the ASTM standard recommendation. The later is therefore not arbitrary but can reflect properly the effects of sample size and geometry.

This limitation due to the sample size exists in nearly all the studies, and especially for bio-inspired structures as it is rather difficult to process samples with large thicknesses. All the studies published so far, including the ones discussed here, were made with samples of very similar thickness, in the 2-3 mm range [2-5,7].

Other limitations exist on these methods to determine the R curve for materials with high energy consumption during fracture, especially regarding the estimation of the crack extension. Extensive crack deviation (anisotropic microstructure, presence of interfaces) raises the question of the use of the real crack length or the projected one. The estimation of the crack length in our work is made with a compliance method [6,8]. The validity of the method was checked by comparing the optically measured actual crack projected length at the end of the test with the value using the method, which were found to be in good agreement (mainly because our sample is fully ceramic i.e., with limited plasticity). The elastic compliance can also be measured by partially unloading the sample [2,8], a method that has also its own limitations: less points can be measured on the R-curve, making the determination of an eventual inflexion point more difficult, and some energy dissipation can occur during the unloading/loading cycles. More precise measurements of the crack length should thus be performed, as well as measuring crack extension at the surface and in the volume of the sample (for example with X-ray computed tomography). The effect of such a large deflection on the relationship compliance/crack length is still to be carefully assessed, and could modify the shape of the calculated R-curve.

We agree that either one of the conservative criteria (the ASTM standard or the approach proposed by Prof. Ritchie) can provide an estimation of the toughness of these samples and to be able to compare these values with existing materials. The ASTM standard has the advantage of being less arbitrary: the determination of an inflexion point in the R-curve can be difficult, in particular when a limited number of data points are obtained. Work is currently underway in our lab to measure the R-curves on much thicker samples. Although the first results support the results discussed here, we will report on this only once a complete and reproducible characterization has been performed. In the meantime, the best practices are probably to compare toughness values measured using the same validity criterion.

We fully agree with Prof. Ritchie that reproducibly processing larger materials is essential and should be a concern for our community. Improvements of the processing routes may provide larger samples, much needed not only to better understand their true properties, but also to pave the way for actual technological applications. Related to this scale up is the need for simple and reproducible processing routes for these materials, if we want strong and tough materials to become a reality and not remain a result of academic interest only, like many studies before.

The objective of this work was to prove the validity of a different approach to the development of strong and tough materials. Instead of relying on the presence of a ductile phase, which always results in moderate level of strength, we demonstrated that strong and tough ceramics can be produced by combining brittle materials in some special bioinspired way. Even by taking the most conservative criteria for the toughness determination, the properties of these ceramics are excellent. Beyond the technical debate on the determination of properties of these materials, we believe that these toughening concepts are already working and that they are worth further exploration. As usual, the future will tell if these materials have a real potential in applications.

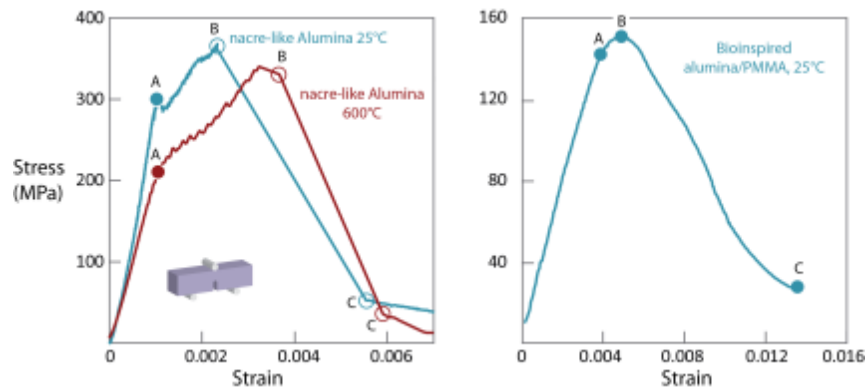

Figure 1: Stress-strain curves for SENB samples of bioinspired ceramic/ceramic composites [3] and ceramic/polymer composites [2]. Different toughening mechanisms are found in both cases, leading to different shapes of the stress-strain curves and thus of the R-curves. The crack starts to propagate in point A. Strain-hardening, with an increase of the stress as the crack propagates, is observed until point B; the strain hardening is much more important in the ceramic/ceramic composite, where a very important crack deflection is observed, while almost no strain hardening can be observed in the ceramic/polymer composite, with little crack deflection. The crack is then unstable from B to C in the ceramic/ceramic composite, while it is stable in the case of the ceramic/polymer composite, with evidences of extensive crack bridging.

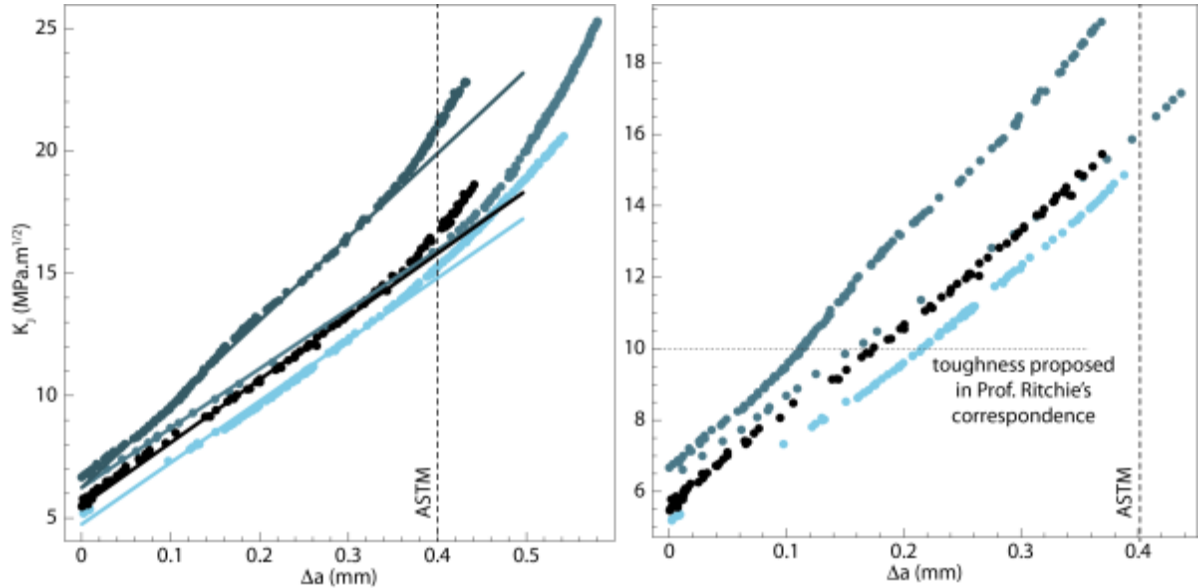

Figure 2: R-curves of nacre-like alumina [3], at room temperature and 600°C. The linear fit of the first part of the curve help identify the location of the inflexion point of the R-curve, as proposed in Prof. Ritchie's correspondence, leading to the crack extension where the toughness can become geometry-dependent. A close up view, where the section of the R-curve after the inflexion point has been removed, is shown on the right hand side. The maximum crack extension based on the ASTM standard is indicated. The maximum valid crack extension proposed in Prof. Ritchie's correspondence is therefore very close to that defined by the ASTM standard. The corresponding toughness at this extension are thus in the 15-19 MPa.m$^{1/2}$ range, with an average value of 16.6 MPa.m$^{1/2}$. The average value of toughness measured using the ASTM standards is 17.3 MPa.m$^{1/2}$